\documentclass
[superscriptaddress,secnumarabic,amssymb,amsmath,nobibnotes,aps,prd,showkeys,nofootinbib,nopacsnumber,onecolumn,10pt]{revtex4}%
\usepackage{graphicx}
\usepackage{bm}
\usepackage{xcolor}
\usepackage{amsmath}
\usepackage{amsfonts}
\usepackage[english]{babel}
\usepackage{amssymb}
\usepackage[utf8]{inputenc}%
\providecommand{\U}[1]{\protect\rule{.1in}{.1in}}

\newcommand{\be}{\begin{equation}}
\newcommand{\ee}{\end{equation}}

\newcommand{\mincir}{\raise
-3.truept\hbox{\rlap{\hbox{$\sim$}}\raise4.truept\hbox{$<$}\ }}
\newcommand{\magcir}{\raise
-3.truept\hbox{\rlap{\hbox{$\sim$}}\raise4.truept\hbox{$>$}\ }}

\begin{document}
\title{Saturation Mechanisms in the Interacting Dark Sector}
\author{Andronikos Paliathanasis}
\email{anpaliat@phys.uoa.gr}
\affiliation{Institute of Systems Science, Durban University of Technology, Durban 4000,
South Africa}
\affiliation{Centre for Space Research, North-West University, Potchefstroom 2520, South Africa}
\affiliation{Departamento de Matem\`{a}ticas, Universidad Cat\`{o}lica del Norte, Avda.
Angamos 0610, Casilla 1280 Antofagasta, Chile}
\affiliation{National Institute for Theoretical and Computational Sciences (NITheCS), South Africa.}
\author{Kevin J. Duffy}
\email{DuffyK@ukzn.ac.za}
\affiliation{Department of Mathematics, School of Agriculture and Science, University of KwaZulu-Natal, 4041, Durban, South Africa}
\affiliation{National Institute for Theoretical and Computational Sciences (NITheCS), South Africa}

\begin{abstract}
We introduce a family of phenomenological cosmological models featuring an interacting dark sector modulated by a sparseness scale parameter, in order to describe the late-time accelerated expansion of the universe. The sparseness scale, inspired by well-established saturation mechanisms in ecology and biology, is introduced in the interaction as a half-saturation constant that bounds the energy exchange between dark matter and dark energy, controls the dynamical behaviour of the physical variables and can prevent the phantom crossing. We consider three nonlinear interacting models, where two of them recover the linear interacting scenarios when the sparsity parameter vanishes. We examine the phase-space of the cosmological field equations by using the Hubble normalization approach. We determine the stationary points and their stability properties in order to reconstruct the asymptotics behaviour of the field equations. Such an analysis allows us to demonstrate the effects of the sparseness scale on the background dynamics. We test the interacting models with observational data. Specifically, we employ Supernovae catalogues, cosmic chronometers, Baryon Acoustic Oscillation measurements from DESI DR2, and redshift-space distortion measurements of the growth of large-scale structure through the $f$ and $f\sigma_8$ observables. The Bayesian analysis suggests that, for two of the three models, a vanishing sparsity parameter is disfavoured at more than the 95\% confidence interval, providing observational support for a nonzero sparseness scale in the dark sector interaction.

\end{abstract}
\date{\today}
\maketitle

\section{Introduction}

Cosmological models that describe energy transfer between the fluids composing
the dark sector have attracted considerable interest, since they offer a
dynamical framework for dark energy \cite{Amendola:1999er,con1,con2,con3,Berger:2006db,delCampo:2008jx,Sharma:2021ayk} which can explain the recent cosmological observations and alleviate cosmological tensions \cite{ht3,ht4,ht5,ht6,ht7,DiValentino:2017iww,DiValentino:2019ffd,Lucca:2021dxo,Simon:2024jmu}. 

Nonzero Interacting terms can be naturally introduced in the gravitational
field equations when one moves to theories beyond General Relativity. Although
General Relativity does not fully satisfy Mach's principle, the latter finds a
natural realization in Brans-Dicke theory \cite{Brans:1961sx,Brans:1962zz}, defined in the Jordan frame \cite{Flanagan:2004bz}, where
a scalar field couples nonminimally to the gravitational action \cite{Kofinas:2015sjz}. The
introduction of this scalar field is essential, as it dynamically determines
the effective gravitational coupling and dynamics. Under a conformal
transformation to the Einstein frame \cite{Postma:2014vaa,GiontiSJ:2023tgx,Tsamparlis:2013aza,Kaiser:2010ps}, the Brans-Dicke theory can be written in
the form of General Relativity with a minimally coupled scalar field. However,
this transformation introduces nongravitational interaction terms between the
scalar field and the matter sector. As a result, effective interacting terms
between different components follows naturally from the geometric
reformulation of the theory.

Alternatively, interacting scenarios can be produced within different
geometric frameworks, such as Weyl Integrable Spacetime (WIS) \cite{salim96,ww1,ww2,ww3}. In this
setting, spacetime is described by a metric tensor, while the affine
connection that defines the auto-parallels corresponds to that of a conformally
related metric. This geometric structure leads to the introduction of
effectively interacting terms in the matter Lagrangian between the matter and
the conformal factor, which relates the two metric tensors. Apart from the
 two considerations above, there are a plethora of studies in the
framework of modified theories of gravity, where a dark matter and dark
energy interaction is introduced \cite{Delsate:2012ky,Csillag:2025gnz,Wazny:2025jth,Sutiono:2024njb,Paliathanasis:2024sle,Balakin:2018vqv,Bojowald:2023djr,Nozari:2008hz,Paliathanasis:2020wjl,Fomin:2017bjb,Ito:2009nk,Paliathanasis:2024gwp,Dutta:2015xha,df7}.

These theoretical considerations motivate the introduction of the Chameleon mechanism \cite{ch1,ch2}. Due to the 
interacting term in the Action Integral between the scalar field and the
matter Lagrangian, there exists a contribution to the mass of the
scalar field from the energy density of the matter source. Hence, the mass of
the scalar field is different in spaces with a high density of matter, such as
compact objects, compared to spaces with a lower mass density. The Chameleon mechanism ensures compatibility with Solar System tests by dynamically screening the scalar field in high-density regions \cite{Hees:2011mu,Yuan:2025twx,Negrelli:2020yps,Burrage:2018dvt}.

However, in order to understand the nature of the interacting dark sector,
there is a plethora of phenomenological models which have been introduced in
the literature, we refer to the recent review  \cite{vanderWesthuizen:2025vcb,vanderWesthuizen:2025mnw,vanderWesthuizen:2025rip}. The most simple models are those where the interacting
function is a linear function of the energy density of the energy densities
of dark matter and/or dark energy \cite{Valiviita:2009nu,Boehmer:2008av}. However, such interacting
models have limitations since they suffer from instabilities in dark
sector perturbations at early times \cite{Valiviita:2008iv}. This
limitation can be circumvented when a nonlinear interaction is introduced \cite{Koshelev:2010umw}. Recent
observational data from the Dark Energy Spectroscopic Instrument (DESI DR2)
have significantly strengthened the case for studying interacting models \cite{Zhu:2025lrk,Zhang:2025dwu,Li:2025muv,Petri:2025swg,SantanaJunior:2024cug,Li:2025ula,Pan:2025qwy,Silva:2025hxw,Yang:2025uyv,Paliathanasis:2026ymi,Zhai:2025hfi,vanderWesthuizen:2025iam,Sahlu:2026bsa,Shah:2026oxn,Benisty:2024lmj,Zhai:2023yny,Bonilla:2021dql,Duan:2026zqj,Califano:2024xzt,Giare:2024ytc,Silva:2025bnn,Sabogal:2025mkp,Sabogal:2024yha}, as
they can account for the late-time acceleration of the universe without
requiring phantom crossing \cite{Guedezounme:2025wav}. At the same time, such models can provide a
mechanism for generating an effective weak pressure component in the dark
matter sector, a feature that appears to be consistent with current
observational data \cite{Carloni:2025jlk,Yang:2025ume}. 

Within the phenomenological framework, two previously proposed nonlinear interacting models
 \cite{cc001,cc002} were shown
to have similar dynamical features to models applied in population dynamics
\cite{2025PDU4701750P}. Motivated by this analogy, we introduce a new family
of interacting dark sector models characterized by the presence of a sparsity
scale. These models incorporate an additional parameter, the sparsity
parameter, which directly influences the cosmological evolution. Specifically,
this new parameter can control the construction of the phase-space so as to
move the position of the stationary points, as well as to prevent the phantom crossing of the dark energy. In what follows, we analyze the impact of this
sparsity scale on the dynamics of the dark sector and confront the models with
late-time background cosmological observations, as well as redshift-space
distortion measurements that probe the growth of matter perturbations. 
The
structure of the analysis is organized as follows.

In Section \ref{sec2} we briefly discuss, within a phenomenological framework,
the interacting dark sector, and we introduce three models characterized by a
sparseness scale parameter $\zeta$. The role of this sparseness scale on the
dynamical evolution of the physical parameters is examined in Section
\ref{sec3}, where we study the properties of the phase-space for the
cosmological field equations by means of the Hubble normalization approach. In
Section \ref{sec4}, we use background data together with the redshift-space
distortion measurements for the growth of matter in order to constrain the
free parameters of the models. Specifically, for the background data we
consider supernovae, cosmic chronometers and baryon acoustic oscillations,
while for the growth of matter perturbations, we employ the $f$ and
$f\sigma_{8}$ data. Furthermore, we perform a statistical comparison of our
models with the $\Lambda$CDM model by applying the Akaike Information
Criterion and the Bayesian evidence. Finally, in Section \ref{sec5} we
summarize our results.

\section{Interacting Dark Sector with a Sparseness Scale}

\label{sec2}

In the standard cosmological scenario, dark matter and dark energy are assumed
to evolve independently and to interact only gravitationally.

Within the framework of a spatially flat and FLRW geometry with line element%
\begin{equation}
ds^{2}=-dt^{2}+a^{2}\left(  t\right)  \left(  dx^{2}+dy^{2}+dz^{2}\right)  ,
\label{fe.01}%
\end{equation}
the cosmological field equations are\footnote{Since our focus is on the
late-time universe, we neglect the contribution of radiation.}
\begin{align}
3H^{2}  &  =\rho_{m}+\rho_{b}+\rho_{d},\label{fe.02}\\
-2\dot{H}-3H^{2}  &  =p_{d}, \label{fe.03}%
\end{align}
in which $\rho_{b}$ is the energy density of the baryons, $\rho_{m}$ is the
pressureless dark matter and $\rho_{d}$ is the dark energy component with
pressure $p_{d}$ and equation of state parameter $w_{d}$, defined by the
expression $p_{d}=w_{d}\rho_{d}$. The last two fluid components form the dark
sector of the universe.

Furthermore, the Bianchi identity reveal the conservation law for the cosmic
fluid
\begin{equation}
\left(  \rho_{m}+\rho_{b}+\rho_{d}\right)  ^{\cdot}+3H\left(  \rho_{m}%
+\rho_{b}+\rho_{d}+p_{d}\right)  =0, \label{fe.04}%
\end{equation}
\bigskip and since the components evolve independently it follows%
\begin{align}
\dot{\rho}_{b}+3H\rho_{b}  &  =0,\label{fe.05}\\
\dot{\rho}_{m}+3H\rho_{m}  &  =0,\label{fe.06}\\
\dot{\rho}_{d}+3H\left(  1+w_{d}\right)  \rho_{d}  &  =0. \label{fe.07}%
\end{align}
from where it follows $\rho_{b}=3H_{0}^{2}\Omega_{b0}a^{-3}$,~$\rho_{m}%
=3H_{0}^{2}\Omega_{m0}a^{-3}\,\ $and $\rho_{d}=3H_{0}^{2}\Omega_{d0}%
\exp\left(  -3\int_{a0}^{a}\frac{1+w_{d}}{\alpha}d\alpha\right)  $.

Therefore, the expression for the background spacetime is%
\begin{equation}
\left(  \frac{H\left(  a\right)  }{H_{0}}\right)  ^{2}=\Omega_{b0}%
a^{-3}+\left\{  \Omega_{m0}a^{-3}+\Omega_{d0}\exp\left(  -3\int_{a_{0}}%
^{a}\frac{1+w_{d}}{\alpha}d\alpha\right)  \right\}  , \label{fe.08}%
\end{equation}
where in the case $w_{d}=-1$, the standard $\Lambda$CDM model is recovered.
The components within the bracket in (\ref{fe.08}) correspond to the dark sector.

From a phenomenological perspective, several frameworks have been proposed in
which dark energy and dark matter interact through a dynamical
nongravitational coupling.~Interacting dark sector models have thus been
extensively considered as alternative cosmological models to the $\Lambda
$CDM.\ These models can describe dynamical dark energy component and may
provide a mechanism to avoid effective phantom behavior.

In the interacting scenarios the Hubble function (\ref{fe.08}) is expressed
as
\begin{equation}
\left(  \frac{H\left(  a\right)  }{H_{0}}\right)  ^{2}=\Omega_{b0}%
a^{-3}+\Omega_{DS}\left(  a\right)  ,
\end{equation}
where $\Omega_{DS}\left(  a\right)  $ is the energy density for the dark
sector, with $\Omega_{DS}\left(  a\right)  =3H_{0}\left(  \rho_{m}+\rho
_{d}\right)  $, and the conserved equations (\ref{fe.06}), (\ref{fe.07}) are
modified as
\begin{align}
\dot{\rho}_{m}+3H\rho_{m}  &  =Q,\\
\dot{\rho}_{d}+3H\left(  1+w_{d}\right)  \rho_{d}  &  =-Q.
\end{align}
The function $Q$ represents the interaction term controlling the energy
exchange between dark energy and dark matter. For a positive function $Q>0$,
there is energy transfer from dark energy to dark matter, while for a negative
function $Q<0$ dark matter is converted into dark energy. The choice of the
interaction function $Q$ determines the dynamical behaviour and physical
properties of the dark sector, and different functions $Q$ affects the
evolution of the universe.

The simplest interacting scenarios are described by linear interaction \cite{Valiviita:2009nu,Boehmer:2008av}
functions
\begin{align}
Q_{1}  &  =3\alpha H\rho_{m},~\\
Q_{2}  &  =3\alpha H\rho_{d},\\
Q_{3}  &  =3\alpha H\left(  \rho_{m}+\rho_{d}\right)  ,
\end{align}
where $\alpha$ is a constant dimensionless parameters. In some recent studies
$\alpha$ is considered to be dynamical allowing the interaction to change sign
or providing a mechanism for its onset.

Nonlinear interacting models have also drawn the attention, such as the \cite{Koshelev:2010umw}%
\begin{align}
Q_{4}  &  =\frac{\alpha}{H}\rho_{m}\rho_{d}\\
Q_{5}  &  =\frac{\alpha}{H}\rho_{m}\rho_{d}+3\beta H\rho_{d}.
\end{align}
\qquad As discussed in \cite{2025PDU4701750P}, these two interacting models share a
similar dynamical structure with dynamical systems arising in population
dynamics. Model $Q_{4}$ has been characterized as the compartmental
interaction, while model $Q_{5}$ is identified as the coexistence model, as it
reproduces dynamics analogous to the Lotka--Volterra system. For a recent
discussion on nonlinear interacting dark energy models, we refer the reader to
\cite{vanderWesthuizen:2025rip}.

\subsection{Sparseness Scale} 

In this work, we introduce the nonlinear interacting models
\begin{align}
Q_{A}  &  =3\alpha H\frac{\rho_{m}\rho_{d}}{3\zeta H^{2}+\rho_{d}},\\
Q_{B}  &  =3\alpha H\frac{\rho_{m}\rho_{d}}{3\zeta H^{2}+\rho_{m}},\\
Q_{C}  &  =\frac{\alpha}{H}\frac{\rho_{m}\rho_{d}^{2}}{3\zeta H^{2}%
+\rho_{d}}%
\end{align}
where $\alpha$ is the coupling strength that drives energy transfer and
the constant $\zeta$ defines the \textquotedblleft sparseness
scale\textquotedblright, analogous to the half-saturation constant in
ecological rate equations. Moreover, when $\zeta=0$, interacting models
$Q_{1}$ $Q_{2}$ and $Q_{3}~$are recovered. In order to avoid singular
behaviours parameter $\zeta$ is considered to be positive.

The sparseness scale functions as a half-saturation parameter that limits the
effective interaction strength between dark matter and dark energy as the
matter density evolves. This structure mirrors long established interaction
laws in ecology and biology. For example, in Michaelis-Menten kinetics of
enzymatic catalysis \cite{siri} and growth kinetics for microbial populations
\cite{monod}. In ecological systems, similar saturating responses  have been identified previously \cite{monod,holling}. For microbial ecology and
predator--prey systems the authors in \cite{may,abrams} \ introduced such saturation terms into the
Lotka-Volterra system to prevent unrealistic indefinite growth or oscillatory
divergence. In a similar framework we consider such a sparseness scale in
cosmology as a phenomenological interacting model, to impose bounds in the
energy exchange and reinforce the stability of the cosmological dynamics.

In what follows, we perform the analysis under the assumption that the dark
energy equation of state parameter $w_{d}$ is constant.

\section{Phase-space Analysis}

\label{sec3}

Within the framework of the Hubble normalization we introduce new
dimensionless dependent variables
\begin{equation}
\Omega_{m}=\frac{\rho_{m}}{3H^{2}}~,~\Omega_{d}=\frac{\rho_{d}}{3H^{2}},\text{
}\Omega_{b}=\frac{\rho_{b}}{3H^{2}},
\end{equation}
and a new independent variable $\tau=\ln a$.

\subsection{Interacting Model $Q_{A}$}

The cosmological field equations for the interacting model $Q_{A}~$are
expressed into the equivalent form%
\begin{align}
\frac{d\Omega_{m}}{d\tau}  &  =\frac{3\Omega_{d}\Omega_{m}}{\zeta+\Omega_{d}%
}\left(  \alpha+w_{d}\left(  \zeta+\Omega_{d}\right)  \right)  ,
\label{in.01}\\
\frac{d\Omega_{d}}{d\tau}  &  =3\Omega_{d}\left(  w_{d}\left(  \Omega
_{d}-1\right)  -\alpha\frac{\Omega_{m}}{\zeta+\Omega_{d}}\right)
,\label{in.02}\\
\frac{d\Omega_{b}}{d\tau}  &  =3w_{d}\Omega_{d}\left(  1-\Omega_{m}-\Omega
_{d}\right)  , \label{in.03}%
\end{align}
with constraint algebraic relation%
\begin{equation}
\Omega_{b}=1-\Omega_{m}-\Omega_{d}. \label{in.04}%
\end{equation}
In the new variables the deceleration parameter $q=-1-\frac{\dot{H}}{H^{2}}$
reads%
\begin{equation}
q=\frac{1}{2}\left(  1+3w_{d}\Omega_{d}\right)  . \label{in.05}%
\end{equation}

\begin{figure}[h]
\centering\includegraphics[width=0.5\textwidth]{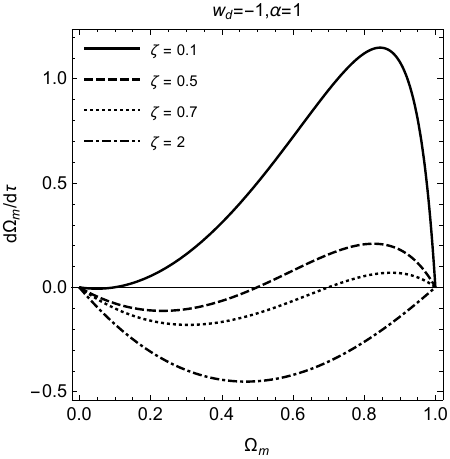}\caption{Interaction
$Q_{A}$: Behaviour of the right-hand side of equation (\ref{in.01}) is shown
for $w_{d}=-1,~\alpha=1$ and various values of $\zeta$ on the surface
$\Omega_{b}\simeq0$. We find that the sparsity scale parameter $\zeta$ plays a
crucial role in determining the sign of the monotonicity of $\Omega_{m}$
within various domains. }%
\label{ss01}%
\end{figure}

To investigate the impact of the sparsity parameter $\zeta$ in the interacting
model, in Fig. \ref{ss01} we present the behaviour of the right-hand side of
equation (\ref{in.01}) for different values of the parameter $\zeta$, on the
asymptotic surface $\Omega_{b}\simeq0$. The sign of this function, which
depends on $\zeta$, determines the energy density rate for dark
energy and dark matter. For positive values of the right-hand side of
equation (\ref{in.01}), $\Omega_{m}$ is an increasing function, while
for negative values, $\Omega_{m}$ decreases. The value of parameter $\zeta$
determines at which point this function changes sign.

We continue our analysis with the determination of the stationary points for
the autonomous dynamical system (\ref{in.01})-(\ref{in.04}). Each stationary
point describes an asymptotic behaviour of the cosmic evolution.

The admitted stationary points $A=\left(  \Omega_{m}\left(  A\right)
,\Omega_{d}\left(  A\right)  ,\Omega_{b}\left(  A\right)  \right)  ~$are%
\begin{align*}
A_{0}  &  =\left(  \Omega_{m},0,1-\Omega_{m}\right)  ,\\
A_{1}  &  =\left(  0,1,0\right)  ,\\
A_{2}  &  =\left( 1+\frac{\alpha}{w_{d}}+\zeta, -\left(  \frac{\alpha}{w_{d}}+\zeta\right) ,0\right)  .
\end{align*}

Point $P_{0}$ describes the\ matter dominated era with deceleration parameter
$q\left(  A_{0}\right)  =\frac{1}{2}$, where there is no interacting
term. On the other hand, point $P_{1}$ describes a universe dominated by
dark energy with $q\left(  A_{1}\right)  =\frac{1}{2}\left(  1+3w_{d}\right)
$. Finally, point $A_{2}$ describes a universe dominated by the dark sector,
with a nonzero interacting component, with $q\left(  A_{2}\right)  =\frac
{1}{2}\left(  1-3\left(  \alpha+\zeta w_{d}\right)  \right)  $, acceleration
occurs when $\left(  \alpha+\zeta w_{d}\right)  >\frac{1}{3}$. In order
to avoid the appearance of ghosts, we introduce the constraint $0\leq-\left(
\frac{\alpha}{w_{d}}+\zeta\right)  \leq1$. The point $A_{2}$ is governed by the
sparsity parameter $\zeta\,\ $and is the point which describes the change
of sign of the evolution equation for $\Omega_{m}$, as described before
and demonstrated in Fig. \ref{ss01}.

Using the constraint (\ref{in.04}), we eliminate $\Omega_{b}$ and reduce the system to the two-dimensional system defined by Eqs.~(\ref{in.01}) and (\ref{in.02}). This formulation allows us to investigate the stability properties of the corresponding stationary points.

For the family of points $A_{0}$ we calculate the eigenvalues $\left\{
0,-3\left(  w_{d}+\frac{\alpha}{\zeta}\Omega_{m}\right)  \right\}  $, where
for $w_{d}+\frac{\alpha}{\zeta}\Omega_{m}<0$, it has a stable submanifold.
Moreover, for point $A_{1}$ we derive the eigenvalues $\left\{
w_{d},3\left(  w_{d}+\frac{\alpha}{1+\zeta}\right)  \right\}  $, and thus for
$w_{d}<0$ and $w_{d}<-\frac{\alpha}{1+\zeta}$, the asymptotic solution is an
attractor.

\ Therefore, for point $A_{2}$ the eigenvalues are $\left\{
-3\left(  \alpha+w_{d}\zeta\right)  ,-\frac{3}{\alpha}\left(  \alpha
+w_{d}\zeta\right)  \left(  w_{d}\left(  1+\zeta\right)  +\alpha\right)
\right\}  ,$ from where WE conclude that when $\left(  \alpha+w_{d}\zeta\right)
>0$ and $\left(  w_{d}\left(  1+\zeta\right)  +\alpha\right)  >0$ the point is
an attractor.

\begin{figure}[h]
\centering\includegraphics[width=1\textwidth]{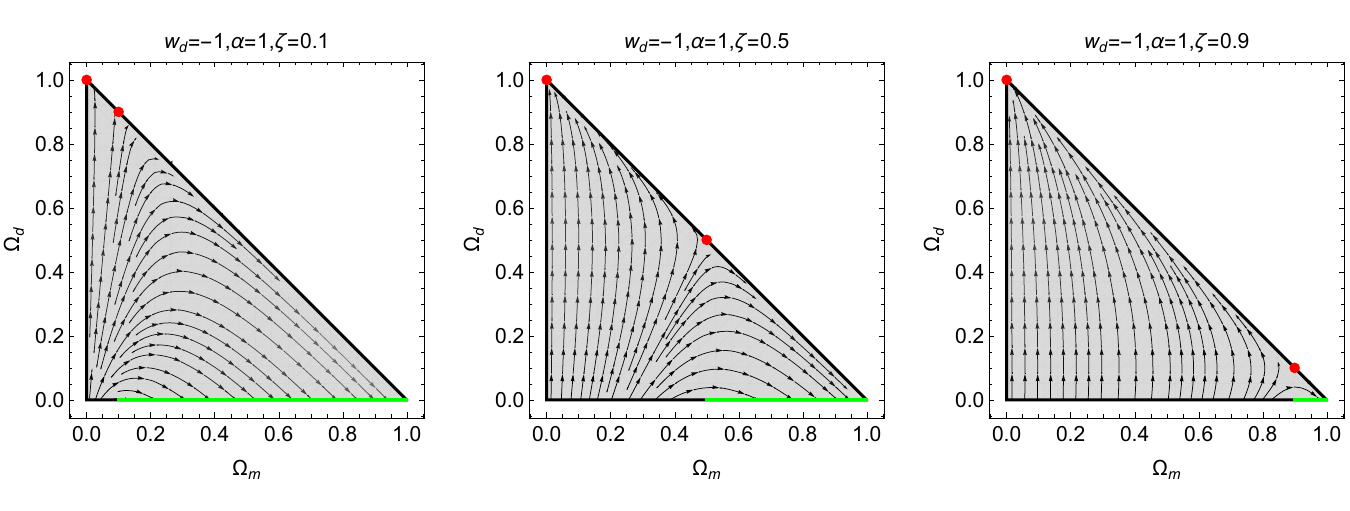}\caption{Interaction
$Q_{A}$: \ Phase-space portraits on the plane $\Omega_{m}-\Omega_{d}$ for the
two-dimensional dynamical system (\ref{in.01}), (\ref{in.02}) for $w_{d}=-1$,
$\alpha=1$ and different values for the sparsity parameter $\zeta$. The red
points correspond to the stationary points $A_{1}$ and $A_{2}$, while the
green line corresponds to the stable submanifold for the family of points
$A_{0}$.}%
\label{ss.02a}%
\end{figure}

In Fig. \ref{ss.02a} we present A phase-space portrait for the two-dimensional
dynamical system (\ref{in.01}), (\ref{in.02}).

\subsection{Interacting Model $Q_{B}$}

For the second interacting model, namely $Q_{B}$, the evolution of
dimensionless variables are described by the dynamical system%
\begin{align}
\frac{d\Omega_{m}}{d\tau} &= \frac{3\Omega_{d}\Omega_{m}}{\zeta+\Omega_{m}}
\left(\alpha + w_{d}\left(\zeta+\Omega_{m}\right)\right), \label{in.06}\\[6pt]
\frac{d\Omega_{d}}{d\tau} &= 3\Omega_{d}\left(w_{d}\left(\Omega_{d}-1\right)
-\alpha\frac{\Omega_{m}}{\zeta+\Omega_{m}}\right), \label{in.07}\\[6pt]
\frac{d\Omega_{b}}{d\tau} &= 3w_{d}\Omega_{d}\left(1-\Omega_{m}-\Omega_{d}\right).
\label{in.08}
\end{align}
with algebraic constraint, expression (\ref{in.04}), and the deceleration
parameter given by expression (\ref{in.05}).

The stationary points $B=\left(  \Omega_{m}\left(  B\right)  ,\Omega
_{d}\left(  B\right)  ,\Omega_{b}\left(  B\right)  \right)  $ are%
\begin{align*}
B_{0}  &  =\left(  \Omega_{m},0,1-\Omega_{m}\right)  ,\\
B_{1}  &  =\left(  0,1,0\right)  ,\\
B_{2}  &  =\left(  1+\frac{\alpha}{w_{d}}+\zeta,-\left(  \frac{w_{d}\zeta}{w_{d}%
}+\zeta\right)  ,0\right).
\end{align*}

The physical properties of the asymptotic solutions at the stationary points
are analogous to those determined before for the interacting model $Q_{A}$.

We linearize the dynamical system around the stationary points, and the family
of points $B_{0}$ have a stable submanifold when $w_{d}+\alpha\frac
{\Omega_{m}}{\zeta+\Omega_{m}}>0$. Moreover, point $B_{1}$ is an attractor
when $w_{d}<0$ and $\alpha+w_{d}\zeta<0$, while $B_{2}$ is an attractor when
$w_{d}\left(  1+\zeta\right)  +\alpha<0$ and $1+\frac{w_{d}\zeta}{\alpha}>0$.

\begin{figure}[h]
\centering\includegraphics[width=0.5\textwidth]{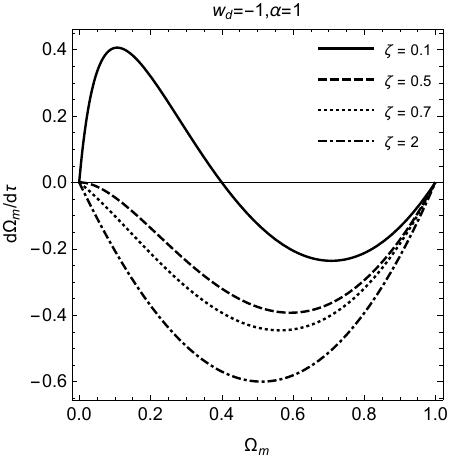}\caption{Interaction
$Q_{B}$: Behaviour of the right-hand side of equation (\ref{in.06}) is shown
for $w_{d}=-1,~\alpha=1$ and various values of $\zeta$ on the surface
$\Omega_{b}\simeq0$. We find that the sparsity scale parameter $\zeta$ plays a
crucial role in determining the sign of the monotonicity of the $\Omega_{m}$
within various domains. }%
\label{ss03}%
\end{figure}

\begin{figure}[h]
\centering\includegraphics[width=1\textwidth]{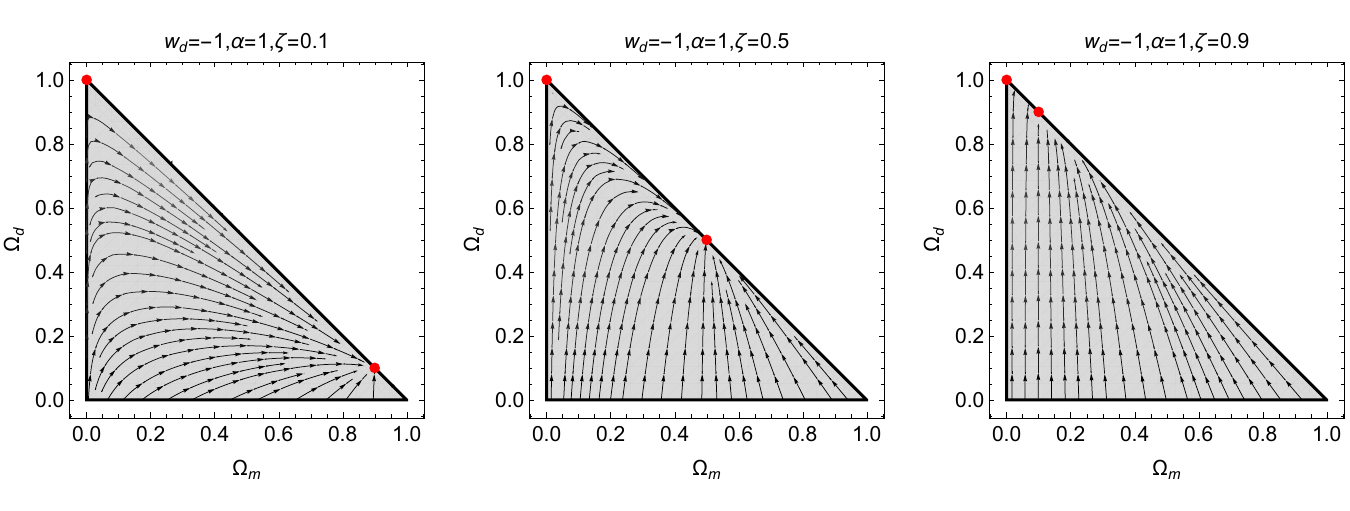}\caption{Interaction
$Q_{B}$: \ Phase-space portraits on the plane $\Omega_{m}-\Omega_{d}$ for the
two-dimensional dynamical system (\ref{in.06}), (\ref{in.07}) for $w_{d}=-1$,
$\alpha=1$ and different values for the sparsity parameter $\zeta$. The red
points correspond to the stationary points $B_{1}$ and $B_{2}$. For the values
of the free parameters used for the plots, the family of points $B_{0}$ are
always sources. }%
\label{ss.04}%
\end{figure}

In Fig. \ref{ss03} we present the behaviour of the right-hand side of equation
(\ref{in.06}) on the surface $\Omega_{b}\simeq0$, from Which we observe the
influence of the sparsity parameter $\zeta\,$ on the dynamics. Furthermore, in
Fig. \ref{ss.04} phase-space portraits of the dynamical system (\ref{in.06}%
)-(\ref{in.08}) are presented.

\subsection{Interacting Model $Q_{C}$}

We continue with the analysis of the phase-space for model $Q_{C}$. The field
equations in the dimensionless variables read%
\begin{align}
\frac{d\Omega_{m}}{d\tau}  &  =\frac{3\Omega_{d}\Omega_{m}}{\zeta+\Omega_{d}%
}\left(  \alpha+w_{d}\left(  \zeta+\Omega_{d}\right)  \right)  ,
\label{ff.01}\\
\frac{d\Omega_{d}}{d\tau}  &  =3\Omega_{d}\left(  w_{d}\left(  \Omega
_{d}-1\right)  -\alpha\frac{\Omega_{d}\Omega_{m}}{\zeta+\Omega_{d}}\right)
,\label{ff.02}\\
\frac{d\Omega_{b}}{d\tau}  &  =3w_{d}\Omega_{d}\left(  1-\Omega_{m}-\Omega
_{d}\right)  ,
\end{align}
with algebraic constraint, expression (\ref{in.04}), and the deceleration
parameter given by expression (\ref{in.05}).

The stationary points $C=\left(  \Omega_{m}\left(  C\right)  ,\Omega
_{d}\left(  C\right)  ,\Omega_{b}\left(  C\right)  \right)  $ are calculated%
\begin{align*}
C_{0}  &  =\left(  \Omega_{m},0,1-\Omega_{m}\right)  ,\\
C_{1}  &  =\left(  0,1,0\right)  ,\\
C_{2}  &  =\left(  1+\frac{w_{d}}{w_{d}+\alpha}\zeta,-\frac{w_{d}}%
{w_{d}+\alpha}\zeta,0\right)  .
\end{align*}
The physical properties of the asymptotic solutions at the stationary points
are analogue to the points found before. However the deceleration parameter
for point $Q_{C}$ is given by the expression $q\left(  Q_{C}\right)  =\frac
{1}{2}\left(  1-3\zeta\frac{w_{d}^{2}}{\alpha+w_{d}}\right)  $. We remark that
point $C_{2}$ exists when $w_{d}+\alpha\neq0$.

From the analysis of the eigenvalues for the linearized system we find that
$C_{0}$ is a source for $w_{d}<0$, point $C_{1}$ is an attractor for $w_{d}<0$
and $\alpha+w_{d}\left(  1+\zeta\right)  <0$, while point $C_{2}$ is an
attractor for $w_{d}+\alpha<0$ and $\frac{w_{d}}{\alpha}\left(  \alpha
+w_{d}\left(  1+\zeta\right)  \right)  <0$.

\begin{figure}[h]
\centering\includegraphics[width=0.5\textwidth]{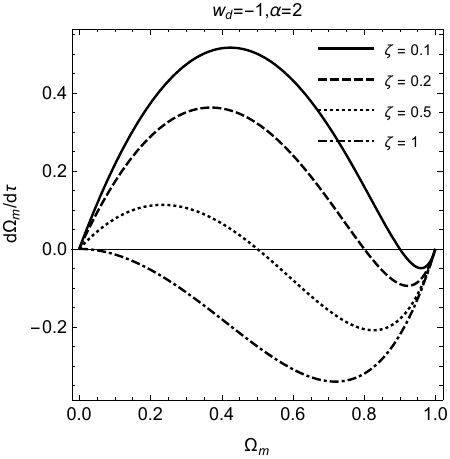}\caption{Interaction
$Q_{C}$: Behaviour of the right-hand side of equation (\ref{ff.01}) is shown
for $w_{d}=-1,~\alpha=2$ and various values of $\zeta$ on the surface
$\Omega_{b}\simeq0$. We find that the sparsity scale parameter $\zeta$ plays a
crucial role in determining the sign of the monotonicity of the $\Omega_{m}$
within various domains.}%
\label{ss.09}%
\end{figure}

\begin{figure}[h]
\centering\includegraphics[width=1\textwidth]{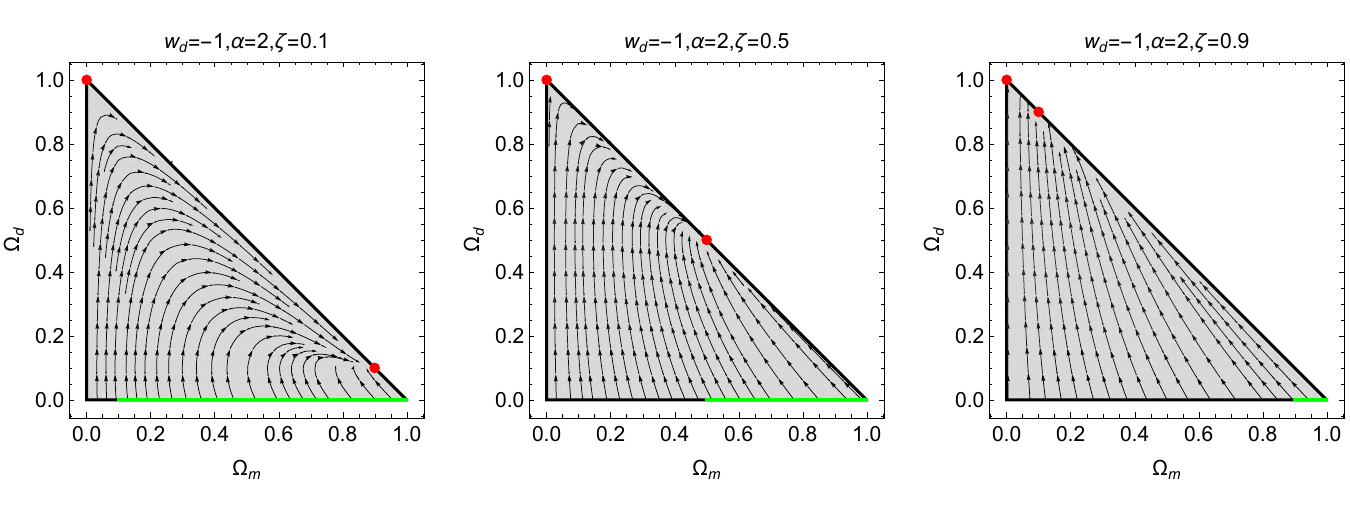}\caption{Interaction
$Q_{C}$: \ Phase-space portraits on the plane $\Omega_{m}-\Omega_{d}$ for the
two-dimensional dynamical system (\ref{ff.01}), (\ref{ff.02}) for $w_{d}=-1$,
$\alpha=2$ and different values for the sparsity parameter $\zeta$. The red
points correspond to the stationary points $C_{1}$ and $C_{2}$. For the values
of the free parameters used for the plots, the family of points $C_{0}$ are
always sources. }%
\label{ss.08}%
\end{figure}

The qualitative evolution of the right-hand side of equation (\ref{ff.01}) on
the surface $\Omega_{b}\simeq0$, are given in Fig. \ref{ss.09}, from where we
observe the effects of the sparsity parameter $\zeta$. Moreover, in Fig.
\ref{ss.08} we present phase-space portraits of the two-dimensional dynamical
system (\ref{ff.01}), (\ref{ff.02}).

\section{Observational Constraints}

\label{sec4}

For our analysis and the comparison of the theoretical predictions with the
cosmological data, we make use of the Bayesian inference framework
Cobaya\footnote{https://cobaya.readthedocs.io/} \cite{cob1,cob2}, with a
custom theory for the derivation of the theoretical observables. For the
sampling algorithm, we employ the PolyChord nested sampler \cite{poly1,poly2},
which makes a direct derivation of the Bayesian evidence $\ln Z$. It
represents the marginal likelihood of the data given a model and it is
obtained by integrating the likelihood function over the full parameter space
weighted by the prior distributions. Moreover, the numerical results are
analyzed with the GetDist library\footnote{https://getdist.readthedocs.io/}%
~\cite{getd}.

In this study, we consider observational data of the late-universe, in
particular we consider the background data provided by the Supernova (SNIa) ,
the cosmic chronometers and the BAO. Specifically, we make use of the
PantheonPlus (PP) \cite{Brout:2022vxf}, Union3.0 (U3) \cite{rubin2023union}
and DES-Dovekie (DESD) \cite{DES:2025sig} SNIa catalogues which late the
luminosity distance at given redshifts. The Hubble direct measurements
presented in\footnote{The likelihood analysis incorporates the full covariance
matrix which can be found in https://gitlab.com/mmoresco/CCcovariance
\cite{Moresco:2020fbm}.} \cite{moresco2020setting} which are derived by the
age difference $\frac{dz}{dt}$ between galaxies at neighboring redshifts, that
is, the cosmic chronometers. Furthermore, for the BAO data we consider the
recent seven measurements for the transverse comoving angular distance ratio,
$D_{M}r_{drag}^{-1},~$the volume-averaged distance ratio, $D_{V}r_{drag}^{-1}$
and and the Hubble distance ratio $D_{H}r_{drag}^{-1}$ provided by the Dark
Energy Spectroscopic Instrument Data Release 2 (DESI DR2)
\cite{DESI:2025zgx,DESI:2025zpo}, where $r_{drag}$ is the sound horizon at the
baryon drag epoch.

In addition, we incorporate the redshift-space distortion (RSD) measurements
that probe the growth of large-scale structure from galaxy redshift surveys
and provide constraints on the matter perturbations. We make use of the data
summarized in \cite{Avila:2022xad,Escobal:2026lnp} and \cite{Avila:2022xad}.
These observables provide measurements of the growth rate of matter
perturbations defined as $f\left(  z\right)  =\frac{d\ln\delta_{m}}{d\ln a}$
and of the $f\sigma_{8}\left(  z\right)  $, where $\sigma_{8}\left(  z\right)
$ represents the root-mean-square amplitude of matter fluctuations in spheres
with radius $8h^{-1}Mpc$, that is, $\sigma_{8}\left(  z\right)  =\sigma
_{8,0}\frac{\delta_{m}\left(  z\right)  }{\delta_{m}\left(  0\right)  }$ and
$\delta_{m}$ describes the evolution of the matter perturbations, that is,
$\delta_{m}=\frac{\Omega_{m}\delta_{dm}+\Omega_{b}\delta_{b}}{\Omega
_{m}+\Omega_{b}}$.\ The perturbation equation for the dark matter $\delta
_{dm}=\frac{\delta\rho_{m}}{\rho_{m}}$ and for the Baryons are presented in
\cite{amendolabook}.

The parameters which are constrained are the Hubble function at the present,
$H_{0}$, the energy density for the dark matter today $\Omega_{m0}$, the sound
horizon at the baryon drag epoch $r_{drag}$, \ the today value of the
root-mean-square amplitude of matter fluctuations in spheres with radius
$8h^{-1}Mpc$, and the two parameters of the interacting models, the sparsity
parameter $\zeta$ and the~$\alpha$, that is, the parametric space has
dimension $\mathcal{N}=6$, that is, $\left\{  H_{0},\Omega_{m0},r_{drag}%
,\alpha,\zeta,\sigma_{8,0}\right\}  $. Moreover, we consider the $\Lambda$CDM
as the base model, where the dimension of the parametric space for the same
dataset is $\mathcal{N}=4$, consisted by the free parameters $\left\{
H_{0},\Omega_{m0},r_{drag},\sigma_{8,0}\right\}  $. \ 

We expect the parameters $\alpha$ and $\zeta$ to be correlated. For this reason, we introduce the redefined parameter $\alpha_{0}$. For model $Q_{A}$, it is defined as 
$\alpha_{0}=3\alpha\frac{\Omega_{d,0}}{\zeta+\Omega_{d,0}}$, 
while for model $Q_{B}$ it becomes 
$\alpha_{0}=3\alpha\frac{\Omega_{d,0}}{\zeta+\Omega_{m,0}}$. 
For model $Q_{C}$, we define 
$\alpha_{0}=\alpha\frac{\Omega_{d,0}^{2}}{\zeta+\Omega_{d,0}}$. Moreover, we compactified the sparsity
parameter $\zeta=\frac{\Delta}{\sqrt{1-\Delta^{2}}}$, with $\Delta\in
\lbrack0,1)$. The limit $\Delta\rightarrow 1$, reaches $\zeta\rightarrow
\infty$.

Moreover, for the present energy density of baryons, we consider
the value obtained by the Planck 2018 collaboration \cite{Planck2018TT},
while for the parameter $w_{d}$ we consider the value $w_{d}=-1$. The priors
considered in the analysis are presented in Table \ref{tabl0}.

%

\begin{table}[tbp] \centering
\caption{Priors of the free parameters.}%
\begin{tabular}
[c]{cc}\hline\hline
\textbf{Parameter} & \textbf{Prior}\\\hline
$H_{0}$ & $\left[  60,80\right]  $\\
$\Omega_{m0}$ & $\left(  0,1\right)  $\\
$r_{drag}$ & $\left[  120,170\right]  $\\
$\alpha_{0}$ & $\left[  -1,1\right]  $\\
$\Delta$ & $[0,1)$\\
$\sigma_{8,0}$ & $\left[  0.3,1.4\right]  $\\\hline\hline
\end{tabular}
\label{tabl0}%
\end{table}%

For the different combinations of the above datasets considered in this work,
we define the maximum value of the $\mathcal{L}_{\max}$ given by the
expression
\[
\mathcal{L}_{\max}^{total}=\mathcal{L}_{\max}^{Data~A}\times\mathcal{L}_{\max
}^{Data~B}\times...,
\]
where $\mathcal{L}_{\max}=\exp\left(  -\frac{1}{2}\chi_{\min}^{2}\right)  $.
Specifically, we consider the following background SNIa+CC+BAO, combined with
the $f\sigma_{8}$, i.e. SNIa+CC+BAO+$f\sigma_{8}$, and the $f+fs_{8}$, i.e.
SNIa+CC+BAO+$f$+$f\sigma_{8},$ where for the SNIa we consider the three
different catalogues. Thus, in total for each interacting model we perform
nine numerical simulations.

We compute Akaike's Information Criterion from the algebraic formula
$AIC=\chi_{\min}^{2}+2\mathcal{N},$ and from the PolyChord sampler we get the
Bayesian evidence $\ln Z$. These two quantities are used to perform a
statistical comparison between models with different numbers of free
parameters $\mathcal{N}$ through the Akaike scale \cite{AIC}, applied to the
difference $\Delta\left(  AIC\right)  =AIC_{1}-AIC_{2}$, and through
Jeffrey's' scale \cite{AIC2} applied to the difference $\Delta\left(  \ln
Z\right)  =\ln Z_{1}-\ln Z_{2}~$. Both criteria penalize models with a larger
number of free parameters, thereby guarding against overfitting.

If $\left\vert \Delta\left(  AIC\right)  \right\vert <2$, Akaike's scale
suggests that the two models are statistically indistinguishable. Moreover, if
$2<\left\vert \Delta\left(  AIC\right)  \right\vert <4$, there is a weak
support for the model with the lower $AIC$ parameter, while for $4 <\left\vert
\Delta\left(  AIC\right)  \right\vert <6$, the support is moderate. Last, but
not least, when $\left\vert \Delta AIC\right\vert >6$ there is clear
evidence in favor of the model with the lower $AIC$. On the other hand, from
Jeffrey's scale it follows that if $\left\vert \Delta\left(  \ln Z\right)
\right\vert <1$, the two models fit the data in a similar way, for
$1<\left\vert \Delta\left(  \ln Z\right)  \right\vert <2.5$ there is weak
evidence in favor of the model with the lower value of $\ln Z$, and if
$2.5<\left\vert \Delta\left(  \ln Z\right)  \right\vert <5$ the evidence is
moderate. Furthermore, for $\left\vert \Delta\left(  \ln Z\right)  \right\vert
>5$, the data provide strong support for the model with the lower value for
the Bayesian evidence.

\subsection{Interacting Model $Q_{A}$}

\begin{figure}[h]
\centering\includegraphics[width=1\textwidth]{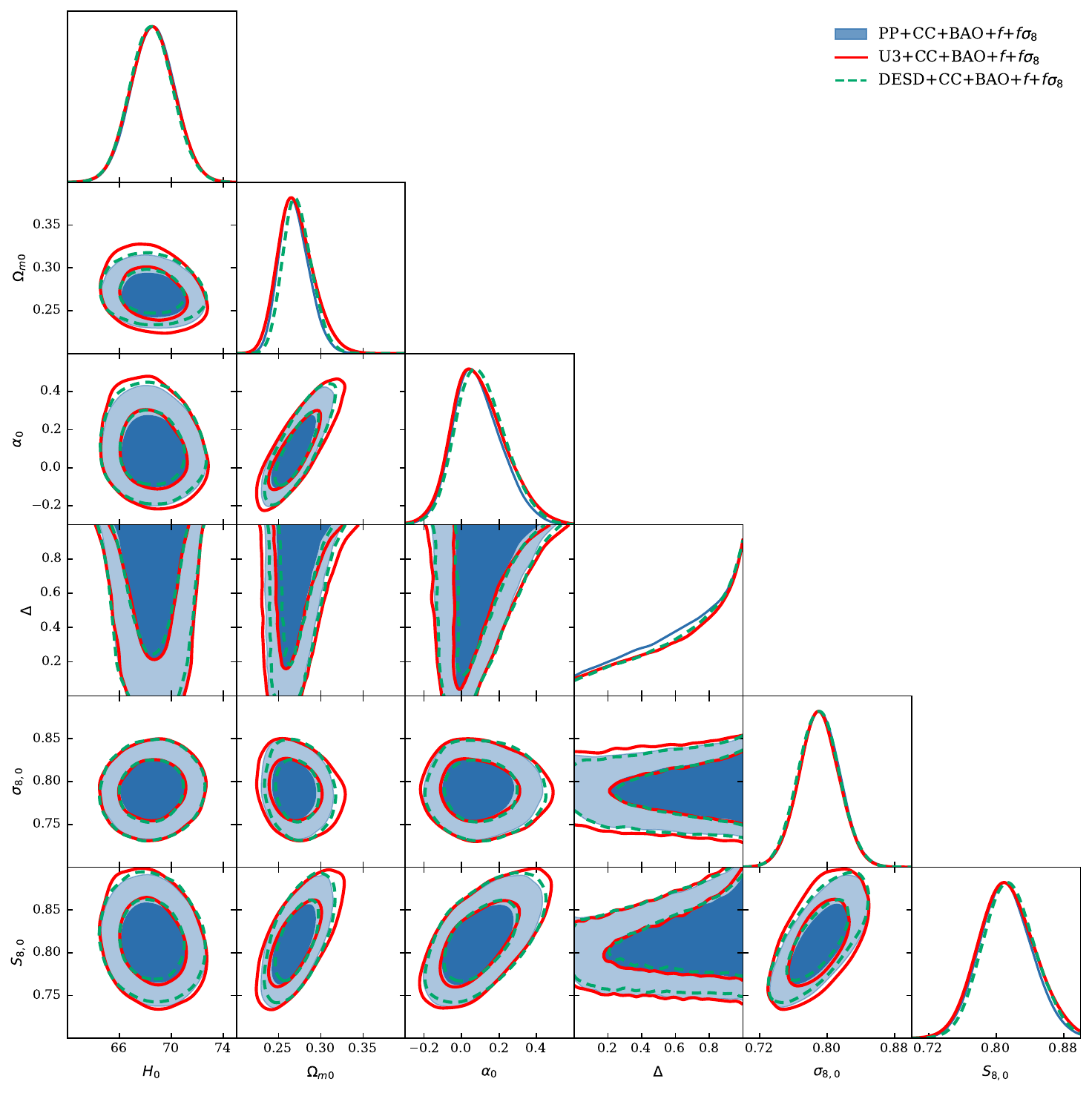}\caption{Interaction
$Q_{A}$: Marginalized posterior contours for the parameters of the interacting
model $Q_{A}$ for the combined datasets. }%
\label{ss05}%
\end{figure}

For the interacting model $Q_{A}$, in Table \ref{tab1} we present the median
values and the marginalized posterior credible intervals (CI) at the $68\%$
level as obtained from the analysis of the numerical outcomes, as also the
comparison of the statistical parameters with that derived for the $\Lambda
$CDM. Moreover, we report the $S_{8,0}=\sigma_{8,0}\sqrt{\frac{\left(
\Omega_{m0}+\Omega_{b0}\right)  }{0.3}}$, which is the normalization of matter
density fluctuations at $8h^{-1}Mpc$. In Fig. \ref{ss05} we present the
marginalized posterior contours for the free parameters of the interacting
model $Q_{A}$.

From the analysis, we observe that the background parameters $H_{0,}%
~\Omega_{m0}$ and $r_{drag}$ are constrained in a similar way, independent from
the SNIa catalogue, with values with mean values $H_{0}\simeq67.4-68.5~\frac
{km}{sMpc}$, $\Omega_{m0}\simeq0.269-0.319$ and $r_{drag}\simeq
147.1-147.2~Mpc$. However, the $\sigma_{8,0}$ is tightly constrained with mean
values $\sigma_{8,0}\simeq0.785-0.791$, and the obtained normalized parameter
for the matter density fluctuations is calculated as $S_{8,0}\simeq0.814-0.824$.
The $S_{8,0}$ obtained are similar to that of the $\Lambda$CDM.
Nevertheless, the $\sigma_{8,0}$ values are slightly lower than that of the
$\Lambda$CDM. The interacting parameter $\alpha_{0}$ is constrained with a
positive preference $a_{0}>0$ within the 68\% CI, when only the background
data are introduced. Nevertheless, when the RSD measurements are introduced,
the $\Lambda$CDM limit, i.e. $\alpha_{0}=0$, is within the 68\% CI. On the
other hand, for the 68\% CI parameter is always $\Delta\succeq0.41$,
indicating that the data support a nonzero sparsity parameter $\zeta$. There
is no upper limit from these datasets, due to the correlation between
$\alpha$ and $\zeta$. Nevertheless, for large values of $\zeta>>1$, parameter
$\alpha_{0}$ indicates the ratio between $\alpha$ and $\zeta$.

The interacting model $Q_{A}$ provides systematic improvement in the maximum
value for the combined likelihood, with lower values for the $\chi_{\min}^{2}%
$, where in comparison to the value derived for the $\Lambda$CDM, $\Delta
\chi_{\min}^{2}$ is within the range $-5.0$ to $-2.1$. However, due to the
different number of degrees of freedom, the value obtained for the $\left\vert
\Delta\left(  AIC\right)  \right\vert $ are within the range $\left[
0.4,1.9\right]  $, from where we conclude that the models are statistically
indistinguishable for all the datasets. However, from Jeffrey's scale for the
Bayesian evidence, since $-2.1\leq\Delta\left(  \ln Z\right)  \leq-1$ for all
the data combinations, excluding the background dataset with the U3 catalogue,
we conclude that the $\Lambda$CDM has weak support from the data over the
interacting model $Q_{A}$.%

\begin{table}[tbp] \centering
\caption{Numerical outcomes for the free parameters for the interacting model $Q_{A}$.}%
\begin{tabular}
[c]{ccccccccccc}\hline\hline
\textbf{Dataset} & $\mathbf{H}_{0}\mathbf{~}$ & $\mathbf{\Omega}_{m0}$ &
$\mathbf{r}_{drag}$ & $\mathbf{\alpha}_{0}$ & $\mathbf{\Delta}$ &
$\mathbf{\sigma}_{8,0}$ & $\mathbf{S}_{8,0}$ & $\Delta\mathbf{\chi}_{\min}%
^{2}$ & $\Delta\left(  \mathbf{AIC}\right)  $ & $\Delta\left(  \ln
\mathbf{Z}\right)  $\\\hline
\textbf{PantheonPlus} &  &  &  &  &  &  &  &  &  & \\\hline
{\small CC+BAO} & $68.0_{-1.7}^{+1.7}$ & $0.292_{-0.021}^{+0.024}$ &
$147.2_{-3.7}^{+3.3}$ & $0.21_{-0.15}^{+0.15}$ & $>0.40$ & $-$ & $-$ & $-3.6$
& $+0.4$ & $-1.1$\\
{\small CC+BAO+}$f\sigma_{8}$ & $68.3_{-1.6}^{+1.6}$ & $0.280_{-0.021}%
^{+0.018}$ & $147.2_{-3.4}^{+3.4}$ & $0.14_{-0.14}^{+0.12}$ & $>0.45$ &
$0.785_{-0.025}^{+0.025}$ & $0.820_{-0.034}^{+0.030}$ & $-2.7$ & $+1.3$ &
$-1.4$\\
{\small CC+BAO+}$f${\small +}$f\sigma_{8}$ & $68.6_{-1.7}^{+1.7}$ &
$0.269_{-0.020}^{+0.015}$ & $147.1_{-3.5}^{+3.5}$ & $0.09_{-0.14}^{+0.11}$ &
$>0.52$ & $0.791_{-0.024}^{+0.024}$ & $0.815_{-0.033}^{+0.029}$ & $-2.7$ &
$+1.3$ & $-1.1$\\\hline
\textbf{Union3} &  &  &  &  &  &  &  &  &  & \\\hline
{\small CC+BAO} & $67.4_{-1.8}^{+1.8}$ & $0.319_{-0.034}^{+0.034}$ &
$147.2_{-3.5}^{+3.5}$ & $0.32_{-0.17}^{+0.17}$ & $>0.43$ & $-$ & $-$ & $-5.0$
& $-1.0$ & $-0.1$\\
{\small CC+BAO+}$f\sigma_{8}$ & $68.0_{-1.7}^{+1.7}$ & $0.291_{-0.030}%
^{+0.025}$ & $147.2_{-3.5}^{+3.5}$ & $0.19_{-0.17}^{+0.15}$ & $>0.49$ &
$0.781_{-0.026}^{+0.026}$ & $0.830_{-0.040}^{+0.035}$ & $-3.4$ & $+0.6$ &
$-1.3$\\
{\small CC+BAO+}$f${\small +}$f\sigma_{8}$ & $68.5_{-1.7}^{+1.7}$ &
$0.271_{-0.024}^{+0.017}$ & $147.1_{-3.4}^{+3.4}$ & $0.10_{-0.16}^{+0.11}$ &
$>0.53$ & $0.790_{-0.024}^{+0.024}$ & $0.814_{-0.036}^{+0.031}$ & $-3.0$ &
$1.0$ & $-2.1$\\\hline
\textbf{DES-D} &  &  &  &  &  &  &  &  &  & \\\hline
{\small CC+BAO} & $68.0_{-1.7}^{+1.7}$ & $0.295_{-0.021}^{+0.021}$ &
$147.2_{-3.5}^{+3.5}$ & $0.23_{-0.14}^{+0.14}$ & $>0.41$ & $-$ & $-$ & $-3.0$
& $+1.0$ & $-1.1$\\
{\small CC+BAO+}$f\sigma_{8}$ & $68.2_{-1.7}^{+1.7}$ & $0.283_{-0.021}%
^{+0.018}$ & $147.2_{-3.5}^{+3.5}$ & $0.16_{-0.15}^{+0.12}$ & $>0.46$ &
$0.784_{-0.026}^{+0.026}$ & $0.824_{-0.033}^{+0.033}$ & $-2.1$ & $+1.9$ &
$-1.0$\\
{\small CC+BAO+}$f${\small +}$f\sigma_{8}$ & $68.5_{-1.7}^{+1.7}$ &
$0.273_{-0.019}^{+0.015}$ & $147.1_{-3.5}^{+3.5}$ & $0.11_{-0.15}^{+0.11}$ &
$>0.55$ & $0.790_{-0.024}^{+0.024}$ & $0.816_{-0.033}^{+0.029}$ & $-2.1$ &
$+1.9$ & $-1.6$\\\hline\hline
\end{tabular}
\label{tab1}%
\end{table}%

\bigskip

\subsection{Interacting Model $Q_{B}$}

\begin{figure}[h]
\centering\includegraphics[width=1\textwidth]{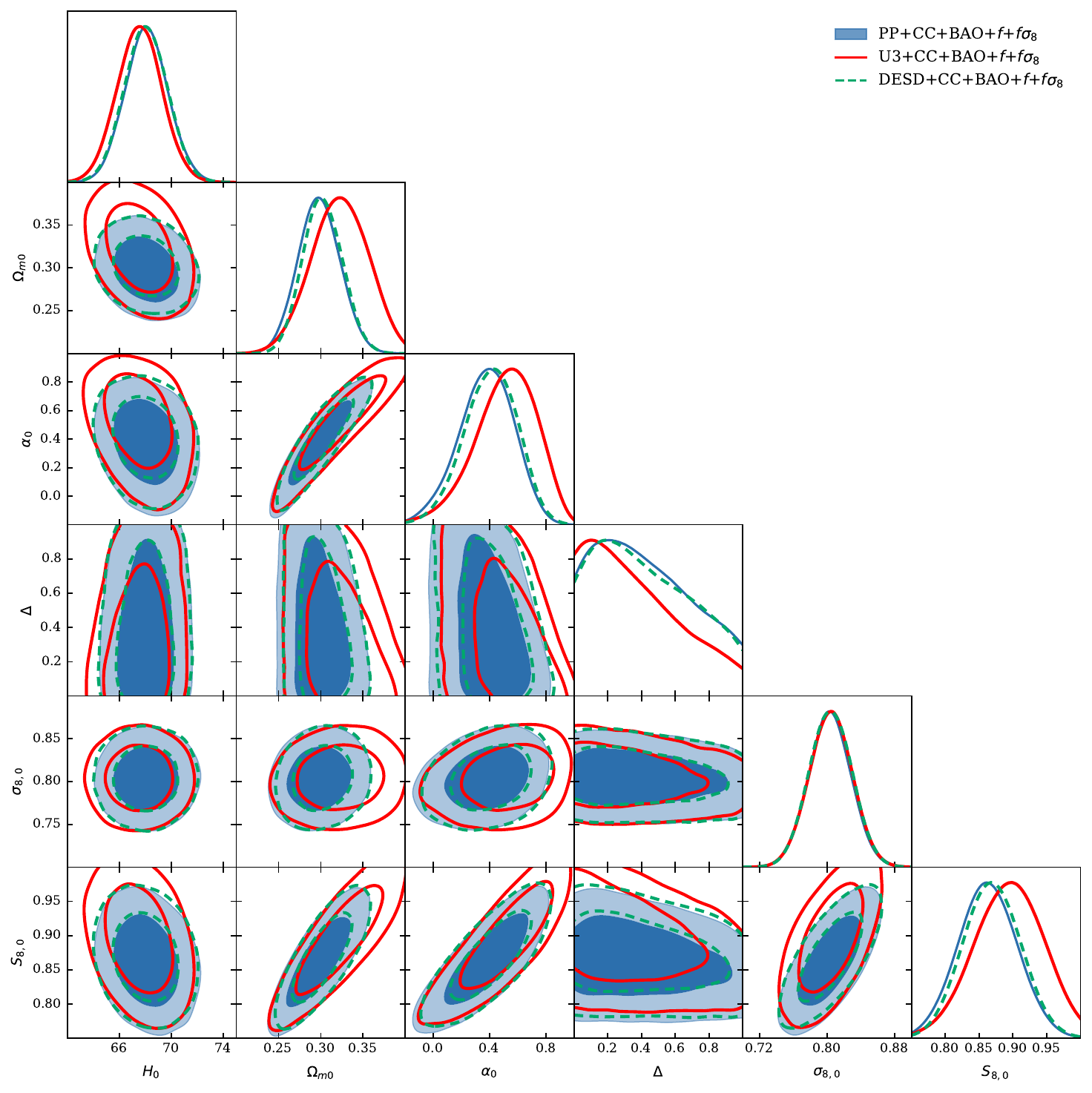}\caption{Interaction
$Q_{B}$: Marginalized posterior contours for the parameters of the interacting
model $Q_{B}$ for the combined datasets. }%
\label{ss06}%
\end{figure}

We continue our discussion with the observational constraints for the
interacting model $Q_{B}$. The mean values for the free parameters and the
marginalized posterior credible intervals at the $68\%$ level are presented in
Table \ref{tab2}. Furthermore, in Fig. \ref{ss06} we present the marginalized
posterior contours.

The constraints provide similar values for the background parameters
$H_{0,}~\Omega_{m0}$ and $r_{drag}$, where now they take values within the
ranges $H_{0}\simeq67.3-68.1~\frac{km}{sMpc}$, $\Omega_{m0}\simeq0.299-0.323$
and $r_{drag}\simeq147.1-147.2~Mpc$. The parameters related to
the growth of matter have mean values within the space $\sigma_{8,0}\simeq0.801-0.804$ and
$S_{8,0}\simeq0.859-0.894$.\ These values are slightly higher than those
obtained from the interacting model $Q_{A}$. The values obtained
for $S_{8,0}$ are larger than that obtained from the weak lensing
cosmological data. The value obtained  for $S_{8,0}$ is
closer to the value obtained by Planck, and this is due to the
larger value for the energy density of dark matter.

From the constraints for the parameters which define the interaction, it follows that the $\Lambda$CDM limit, i.e.$~\alpha_{0}=0$, is
outside the $68\%$ CI for the data combinations considered, indicating some statistical support from the data for the interacting scenario. On the other hand, parameter $\Delta$ has
only an upper limit, $\Delta\leq 0.43$, which means that the value
$\zeta\rightarrow0$ for the sparsity parameter is within~the $68\%$ CI.

As far as the statistical parameters are concerned, the interacting model
$Q_{B}$ provides systematic lower values for the $\chi_{\min}^{2}$, with
$\Delta\chi_{\min}^{2}$ negative for all datasets and in the range
$-6.7<\Delta\chi_{\min}^{2}<-3.5$. The lower values are obtained for the
combined data with the $U3$ catalogue and the $\Delta\chi_{\min}^{2}$ for the
combined data U3+CC+BAO{\small +}$f${\small +}$f\sigma_{8}$. Thus, for this
specific dataset, Akaike's scale suggests weak support for
the interacting model $Q_{B}$. For the rest of the data combinations the
$\left\vert \Delta\left(  AIC\right)  \right\vert $ takes values in the range
$\left[  0.1,1.7\right]  $ indicating that the two models are statistically
equivalent. Nevertheless, from the value of the Bayesian evidence, Jeffrey's
scale indicates that the models are statistically indistinguishable for all the datasets.%

\begin{table}[tbp] \centering
\caption{Numerical outcomes for the free parameters for the interacting model $Q_{B}$.}%
\begin{tabular}
[c]{ccccccccccc}\hline\hline
\textbf{Dataset} & $\mathbf{H}_{0}\mathbf{~}$ & $\mathbf{\Omega}_{m0}$ &
$\mathbf{r}_{drag}$ & $\mathbf{\alpha}_{0}$ & $\mathbf{\Delta}$ &
$\mathbf{\sigma}_{8,0}$ & $\mathbf{S}_{8,0}$ & $\Delta\mathbf{\chi}_{\min}%
^{2}$ & $\Delta\left(  \mathbf{AIC}\right)  $ & $\Delta\left(  \ln
\mathbf{Z}\right)  $\\\hline
\textbf{PantheonPlus} &  &  &  &  &  &  &  &  &  & \\\hline
{\small CC+BAO} & $68.1_{-1.7}^{+1.7}$ & $0.299_{-0.027}^{+0.027}$ &
$147.1_{-3.5}^{+3.5}$ & $0.35_{-0.19}^{+0.25}$ & $<0.62$ & $-$ & $-$ & $-3.9$
& $+0.1$ & $-0.4$\\
{\small CC+BAO+}$f\sigma_{8}$ & $68.1_{-1.7}^{+1.7}$ & $0.297_{-0.026}%
^{+0.026}$ & $147.2_{-3.5}^{+3.5}$ & $0.34_{-0.19}^{+0.24}$ & $<0.59$ &
$0.801$ & $0.859_{-0.046}^{+0.046}$ & $-3.9$ & $+0.1$ & $-0.1$\\
{\small CC+BAO+}$f${\small +}$f\sigma_{8}$ & $68.1_{-1.7}^{+1.7}$ &
$0.298_{-0.025}^{+0.025}$ & $147.3_{-3.4}^{+3.4}$ & $0.37_{-0.18}^{+0.22}$ &
$<0.43$ & $0.803_{-0.025}^{+0.025}$ & $0.863_{-0.044}^{+0.044}$ & $-4.9$ &
$-0.9$ & $+1.0$\\\hline
\textbf{Union3} &  &  &  &  &  &  &  &  &  & \\\hline
{\small CC+BAO} & $67.3_{-1.8}^{+1.8}$ & $0.331_{-0.038}^{+0.038}$ &
$147.2_{-3.5}^{+3..5}$ & $0.52_{-0.17}^{+0.27}$ & $<0.62$ & $-$ & $-$ & $-5.7$
& $-1.7$ & $+0.7$\\
{\small CC+BAO+}$f\sigma_{8}$ & $67.4_{-1.8}^{+1.8}$ & $0.324_{-0.036}%
^{+0.036}$ & $147.2_{-3.5}^{+3.5}$ & $0.49_{-0.18}^{+0.28}$ & $<0.55$ &
$0.801_{-0.026}^{+0.026}$ & $0.893_{-0.050}^{+0.057}$ & $-5.7$ & $-1.7$ &
$+0.5$\\
{\small CC+BAO+}$f${\small +}$f\sigma_{8}$ & $67.5_{-1.7}^{+1.7}$ &
$0.323_{-0.034}^{+0.034}$ & $147.3_{-3.5}^{+3.2}$ & $0.51_{-0.18}^{+0.25}$ &
$<0.48$ & $0.804_{-0.025}^{+0.025}$ & $0.894_{-0.052}^{+0.052}$ & $-6.7$ &
$-2.7$ & $+0.7$\\\hline
\textbf{DES-D} &  &  &  &  &  &  &  &  &  & \\\hline
{\small CC+BAO} & $68.0_{-1.7}^{+1.7}$ & $0.302_{-0.024}^{+0.024}$ &
$147.2_{-3.5}^{+3.5}$ & $0.28_{-0.18}^{+0.23}$ & $<0.63$ & $-$ & $-$ & $-3.5$
& $+0.5$ & $-0.3$\\
{\small CC+BAO+}$f\sigma_{8}$ & $68.0_{-1.7}^{+1.7}$ & $0.301_{-0.024}%
^{+0.024}$ & $147.2_{-3.5}^{+3.5}$ & $0.38_{-0.18}^{+0.23}$ & $<0.59$ &
$0.802_{-0.026}^{+0.026}$ & $0.866_{-0.045}^{+0.045}$ & $-3.5$ & $+0.5$ &
$+0.5$\\
{\small CC+BAO+}$f${\small +}$f\sigma_{8}$ & $68.0_{-1.7}^{+1.7}$ &
$0.302_{-0.023}^{+0.023}$ & $147.2_{-3.2}^{+3.2}$ & $0.40_{-0.17}^{+0.21}$ &
$<0.56$ & $0.804_{-0.025}^{+0.025}$ & $0.869_{-0.043}^{+0.043}$ & $-4.6$ &
$-0.6$ & $+0.9$\\\hline\hline
\end{tabular}
\label{tab2}%
\end{table}%

\subsection{Interacting Model $Q_{C}$}

We continue our discussion with the analysis of the numerical chains for the
model $Q_{C}$. The constraints provide similar values for the background
parameters $H_{0,}~\Omega_{m0}$ and $r_{drag}$ , where now they take mean
values within the ranges $H_{0}\simeq67.2-68.3~\frac{km}{sMpc}$, $\Omega
_{m0}\simeq0.297-0.324$ and $r_{drag}\simeq147.1-147.2~Mpc$. However,
parameter $\alpha_{0}$ is always positive with the $68\%$ CI, for all the
combined data sets, indicating that the $\Lambda$CDM is not within this CI.
Parameter $\Delta$ has a lower bound, which indicating a preference of the
nonzero sparsity parameter $\zeta$. As far as the $\sigma_{8,0}$ and $S_{8,0}$
are concerned, they are found to be similar with that obtained from the $Q_{B}$
model. Specifically the mean values are within the ranges $\sigma_{8,0}%
\simeq0.804-0.810$ and $S_{8,0}\simeq0.862-0.903$. The large values of
$S_{8,0}$ are due to the large  value obtained  for the $\Omega_{m0}$.

Furthermore, from the values of the $\Delta\chi_{\min}^{2}$, $\Delta\left(
AIC\right)  $ and $\Delta\left(  \ln Z\right)  $ we conclude that the
interacting model $Q_{C}$ provides systematically smaller values for the
$\chi_{\min}^{2}$, for all the data sets, not only in comparison to the
$\Lambda$CDM but to the other two interacting models. The difference becomes
larger when the $f+f\sigma_{8}$ data are introduced to the background. From
Akaike's scale we infer that for the combined data U3+CC+BAO and
U3+CC+BAO+$f$+$f\sigma_{8}$, there is weak evidence in favor of the
interacting model. However, the Bayesian evidence reveals that the two models
fit the cosmological data in a similar way.

\begin{figure}[h]
\centering\includegraphics[width=1\textwidth]{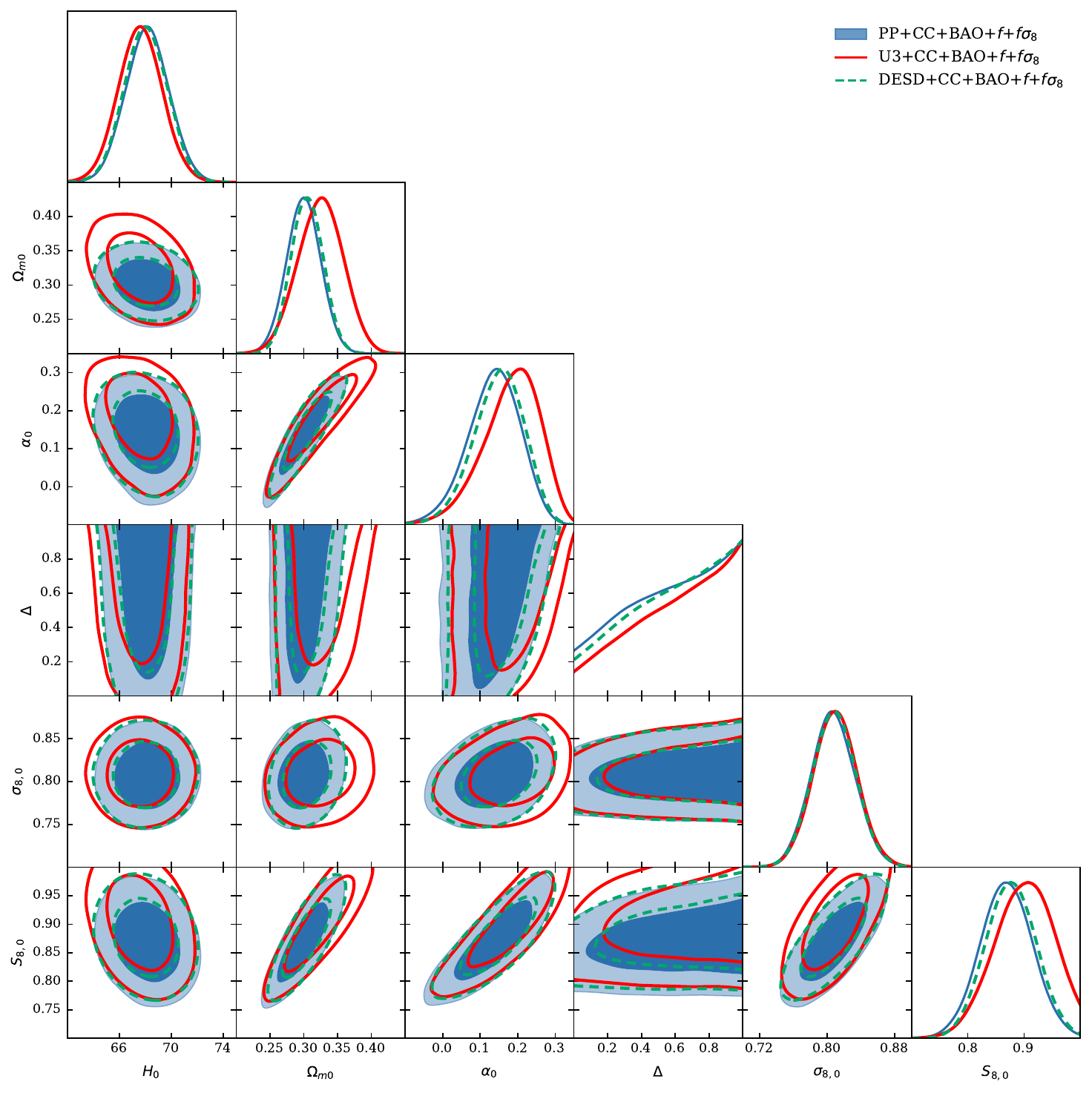}\caption{Interaction
$Q_{C}$: Marginalized posterior contours for the parameters of the interacting
model $Q_{C}$ for the combined datasets. }%
\label{ss07}%
\end{figure}

The above results are summarized in Table \ref{tab3}, while in Fig. \ref{ss07} we
give the marginalized posterior contours.%

\begin{table}[tbp] \centering
\caption{Numerical outcomes for the free parameters for the interacting model $Q_{C}$.}%
\begin{tabular}
[c]{ccccccccccc}\hline\hline
\textbf{Dataset} & $\mathbf{H}_{0}\mathbf{~}$ & $\mathbf{\Omega}_{m0}$ &
$\mathbf{r}_{drag}$ & $\mathbf{\alpha}_{0}$ & $\mathbf{\Delta}$ &
$\mathbf{\sigma}_{8,0}$ & $\mathbf{S}_{8,0}$ & $\Delta\mathbf{\chi}_{\min}%
^{2}$ & $\Delta\left(  \mathbf{AIC}\right)  $ & $\Delta\left(  \ln
\mathbf{Z}\right)  $\\\hline
\textbf{PantheonPlus} &  &  &  &  &  &  &  &  &  & \\\hline
{\small CC+BAO} & $68.0_{-1.7}^{+1.7}$ & $0.301_{-0.027}^{+0.027}$ &
$147.2_{-3.4}^{+3.4}$ & $0.13_{0.066}^{+0.087}$ & $>0.38$ & $-$ & $-$ & $-4.0$
& $+0.0$ & $-0.1$\\
{\small CC+BAO+}$f\sigma_{8}$ & $68.1_{-1.7}^{+1.7}$ & $0.297_{-0.026}%
^{+0.026}$ & $147.1_{-3.5}^{+3.5}$ & $0.12_{-0.067}^{+0.085}$ & $>0.40$ &
$0.804_{-0.026}^{+0.026}$ & $0.862_{-0.048}^{+0.048}$ & $-3.9$ & $+0.1$ &
$-0.5$\\
{\small CC+BAO+}$f${\small +}$f\sigma_{8}$ & $68.2_{-1.7}^{+1.7}$ &
$0.300_{-0.025}^{+0.025}$ & $147.2_{-3.5}^{+3.5}$ & $0.14_{-0.062}^{+0.077}$ &
$>0.43$ & $0.807_{-0.026}^{+0.026}$ & $0.869_{-0.047}^{+0.047}$ & $-5.1$ &
$-1.1$ & $+0.9$\\\hline
\textbf{Union3} &  &  &  &  &  &  &  &  &  & \\\hline
{\small CC+BAO} & $67.2_{-1.8}^{+1.8}$ & $0.338_{-0.040}^{+0.040}$ &
$147.2_{-3.5}^{+3.5}$ & $0.20_{-0.062}^{+0.099}$ & $>0.39$ & $-$ & $-$ &
$-6.2$ & $-2.2$ & $+0.5$\\
{\small CC+BAO+}$f\sigma_{8}$ & $67.6_{-1.8}^{+1.8}$ & $0.322_{-0.036}%
^{+0.036}$ & $147.2_{-3.5}^{+3.5}$ & $0.17_{-0.063}^{+0.100}$ & $>0.43$ &
$0.807_{-0.026}^{+0.026}$ & $0.897_{-0.058}^{+0.058}$ & $-5.6$ & $-1.6$ &
$+0.0$\\
{\small CC+BAO+}$f${\small +}$f\sigma_{8}$ & $67.6_{-1.7}^{+1.7}$ &
$0.324_{-0.033}^{+0.033}$ & $147.2_{-3.4}^{+3.4}$ & $0.18_{-0.058}^{+0.089}$ &
$>0.49$ & $0.810_{-0.026}^{+0.026}$ & $0.903_{-0.055}^{+0.055}$ & $-7.1$ &
$-3.1$ & $+0.4$\\\hline
\textbf{DES-D} &  &  &  &  &  &  &  &  &  & \\\hline
{\small CC+BAO} & $67.9_{-1.7}^{+1.7}$ & $0.306_{-0.026}^{+0.026}$ &
$147.2_{-3.5}^{+3.5}$ & $0.14_{-0.064}^{+0.082}$ & $>0.39$ & $-$ & $-$ &
$-3.7$ & $+0.3$ & $-0.7$\\
{\small CC+BAO+}$f\sigma_{8}$ & $68.0_{-1.7}^{+1.7}$ & $0.301_{-0.025}%
^{+0.025}$ & $147.3_{-3.5}^{+3.5}$ & $0.13_{-0.064}^{+0.084}$ & $>0.41$ &
$0.805_{-0.027}^{+0.027}$ & $0.869_{-0.048}^{+0.048}$ & $-3.5$ & $+0.5$ &
$+0.0$\\
{\small CC+BAO+}$f${\small +}$f\sigma_{8}$ & $68.0_{-1.7}^{+1.7}$ &
$0.305_{-0.024}^{+0.024}$ & $147.3_{-3.5}^{+3.5}$ & $0.15_{-0.060}^{+0.074}$ &
$>0.46$ & $0.808_{-0.026}^{+.0.026}$ & $0.877_{-0.045}^{+0.045}$ & $-5.2$ &
$-1.2$ & $+0.7$\\\hline\hline
\end{tabular}
\label{tab3}%
\end{table}%

\section{Conclusions}

\label{sec5}

In this work we introduced cosmological models which describe an interacting dark sector with a saturation mechanism, which controls the energy exchange between dark matter and dark energy. Specifically, we introduced three nonlinear interacting models, namely $Q_{A}$, $Q_{B}$ and $Q_{C}$ which depend on the new sparseness scale parameter $\zeta$. In the limit where $\zeta \rightarrow{0}$, models $Q_{A}$ and $Q_{B}$ reduce to linear interacting functions. The sparseness scale controls the energy transfer by introducing bounds on the coupling strength. 

In order to explore the effects of the parameter $\zeta$
on the background dynamics, we presented a detailed phase-space analysis of the cosmological field equations using the Hubble normalization variables. We found that the sparseness scale parameter plays an important role in the structure of the phase-space and can be used to eliminate unphysical solutions admitted by the interacting models.

Furthermore, we employed cosmological data to constrain these models as candidates for the description of dynamical dark energy. We considered combinations of background data with the $f$ and $f\sigma_8$ measurements of the growth of matter. The background data considered in this work consist of SNIa observations, direct Hubble measurements from cosmic chronometers, and BAO data from DESI DR2.

For models $Q_{A}$, and $Q_{C}$ the analysis of the data suggests that a nonzero value of the sparseness scale $\zeta$ is favoured within the 95\% CI for all the data combinations. This is not the case for model $Q_{B}$, where $\zeta=0$ appears to be a preferred value. From the $f$ and $f\sigma_8$ we were able to constrain the $\sigma_8$ and $S_8$ observables. The obtained values for the $\sigma_8$ are consistent with those obtained for the $\Lambda$CDM for all the combined datasets. However, this is not the case for the $S_8$ values. Model $Q_{A}$, provides values of $S_8$ consistent with $\Lambda$CDM, whereas models $Q_{B}$ and $Q_{C}$ provide larger values, closer to those obtained from Planck 2018. The reason for this are the larger values obtained for the $\Omega_{m0}$ for these two models. It is important to note that the dark sector is treated as a single fluid, which we interpret as two distinct components, namely dark matter and dark energy. The $S_8$ characterizes the amplitude of large-scale structure growth at late times. Parameter $\Omega_{m0}$ is an integration constant of the cosmological field equations which we interpret as the energy density of the dark matter; however, in an interacting scenario it is not necessarily the case that all of this component contributes to the growth of matter. 

\begin{figure}[h]
\centering\includegraphics[width=1\textwidth]{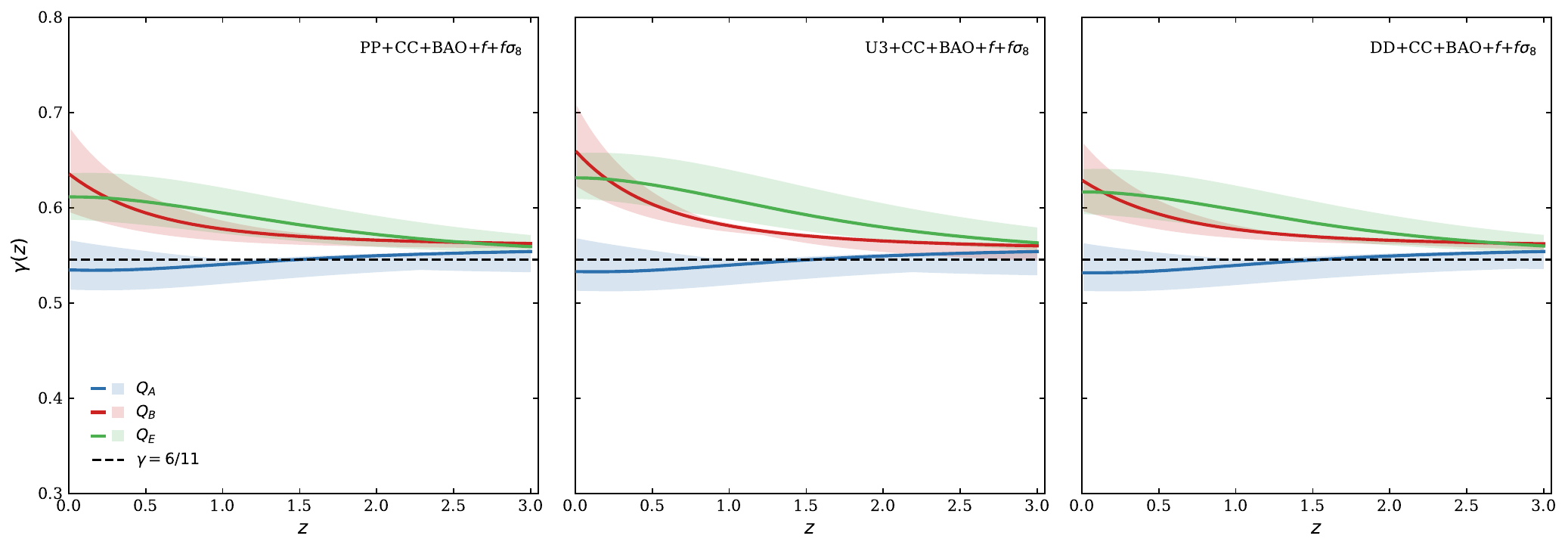}\caption{Evolution of
the growth of index parameter $\gamma\left(  z\right)  =\frac{\ln f\left(
z\right)  }{\ln\Omega\left(  z\right)  }$ for the three interacting models
$Q_{A}$,~$Q_{B}$ and $Q_{C}$ and variables with the 68\% bands as presented in
Tables \ref{tab1}, \ref{tab2} and \ref{tab3} respectively and comparison with
the $\Lambda$CDM value $\gamma_{\Lambda\text{CDM}}=\frac{6}{11}$. }%
\label{ss10}%
\end{figure}

In Fig. \ref{ss10} we present the qualitative evolution of the growth index parameter $\gamma\left( z\right) =\frac{\ln f\left( z\right) }{\ln\Omega\left( z\right) }$ obtained from the posterior distributions of the free parameters from the above observational constraints. We observe that model $Q_{A}$ provides values smaller than the $\Lambda$CDM prediction, that is, $\gamma_{\Lambda}=\frac{6}{11}$, while for $Q_{B}$ and $Q_{C}$ we observe a different behaviour. It is worth noting that in the derivation of the perturbation equations, baryons are assumed to interact with the remaining fluids through gravity only.

From the above analysis we conclude that late-time cosmological data may support interacting dark sector models with a saturation mechanism. In a future work we plan to extend this study to the perturbation level, in order to investigate whether the models are free of instabilities and whether they can describe CMB data.

\begin{acknowledgments}
AP acknowledges the support from FONDECYT Grant 1240514 and from VRIDT through
Resoluci\'{o}n VRIDT No. 096/2022 and Resoluci\'{o}n VRIDT No. 098/2022.
\end{acknowledgments}

\bibliography{biblio}

\end{document}